\documentclass[12pt]{iopart}
\usepackage{graphicx}
\usepackage{url}
\usepackage{amsfonts}
\usepackage{amssymb}

\newcommand{\R}{\mathrm{R}}
\newcommand{\I}{\mathrm{I}}

\begin{document}

\title{Acoustic resonance spectroscopy for the advanced undergraduate laboratory}

\author{J A Franco-Villafa\~ne$^1$, E Flores-Olmedo$^2$, G B\'aez$^2$, O Gandarilla-Carrillo$^3$ and R. A. M\'endez-S\'anchez$^1$}
\address{$^1$ Instituto de Ciencias F\'isicas, Universidad Nacional Aut\'onoma de M\'exico, P. O. Box 48-3, 62251 Cuernavaca, Morelos, M\'exico}
\address{$^2$ Departamento de Ciencias B\'asicas, Universidad Aut\'onoma Metropolitana-Azcapotzalco, Av. San Pablo 180, Col. Reynosa Tamaulipas, 02200 M\'exico D. F., M\'exico}
\address{$^3$ Departamento de Ingenier\'ia El\'ectrica, Universidad Aut\'onoma Metropolitana-Iztapalapa, A. P. 55-534, 09340 M\'exico D. F., Mexico}
\ead{mendez@fis.unam.mx}

\begin{abstract}

We present a simple experiment that allows advanced undergraduates to learn the principles and applications of spectroscopy. The technique, known as {\em acoustic resonance spectroscopy}, is applied to study a vibrating rod. The setup includes electromagnetic-acoustic transducers, an audio amplifier and a vector network analyzer. Typical results of compressional, torsional and bending waves are analyzed and compared with analytical results.
\end{abstract}

\submitto{\EJP}

\maketitle
\section{Introduction}\label{Sec:Intro}

Spectroscopy is a broad field experimental technique in physics with many applications. Historically the spectroscopy was crucial to develop Quantum Mechanics, one of the fundamental pillars of physics. Spectroscopy nowadays is a tool used in research laboratories across the world in Physics, Chemistry and Biology so, it deserves to be introduced to students of physics at an advanced stage of their education. For this reason we have designed an experiment, for an advanced undergraduate --third year-- physics laboratory, based on the acoustic resonance spectroscopy (ARS). This spectroscopy involves scattering of acoustical and mechanical waves; the former for longitudinal (pressure) waves and the latter also includes transverse waves, among others. Compared with other spectroscopic techniques, the ARS is a non-destructive technique that requires minimum sample preparation. The sample used here is a uniform rod of circular cross section that 
can be easily changed by beams or plates uniform or structured with 
some specific purpose. Another advantage is that the ARS results can be analyzed deeply and compared with theoretical predictions. 

In this paper the acoustic resonance spectroscopy is presented with a very simple system: a vibrating rod. In the next section we present the resonant response theory of torsional waves in a rod. In Section~\ref{Sec:Experimental} we present the ARS as well as the experimental setup with a description of the transducers used. A comparison between theory and experiment is performed in Sec.~\ref{Sec:Comparison}. A brief conclusion follows.

\section{Theory: Resonances with losses for torsional waves in rods}
\label{Sec:Theory}

The vibrating rod at low frequencies is one of the simplest cases to study elastic systems. In this regime the elastic rods can vibrate in three different ways: compressional, torsional and bending~\cite{Morietal,Graff,Rossing}; an illustrative animation of these kind of vibrations can be found in~\cite{MendezWeb}. At low amplitudes, and for rods with circular cross section, it is possible to study separately those different kinds of waves. In what follows we will present the theory for torsional vibrations in rods that satisfy the wave equation. The compressional waves, in a first approximation, satisfy also the wave equation while bending (also called flexural) waves satisfy a fourth order partial differential equation~\cite{Graff,Landau1}. The theory developed below can be easily adapted, without major effort, to compressional and bending waves.

\begin{figure}[!ht]
\begin{center}
 \includegraphics[width=0.6\columnwidth]{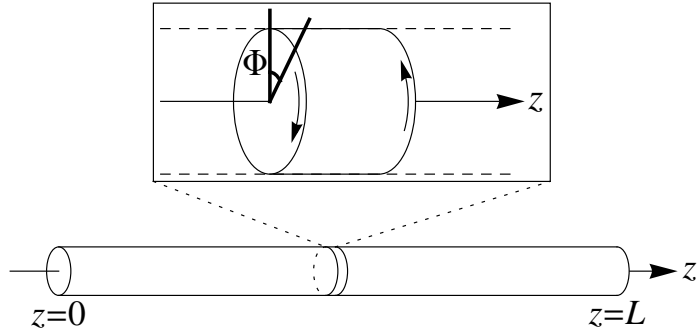}
\end{center}
\caption{Uniform rod with circular cross-section. In the zoom the angle of twist $\Phi$ is defined. The arrows indicate the torque.}
\label{Fig:torsionalwaves} 
\end{figure}

The torsional vibrations in rods with circular uniform cross-section (see figure~\ref{Fig:torsionalwaves}) satisfy
\begin{equation}\label{Eq:linearwaves}
\frac{\partial ^2 \Phi}{\partial z^2}-\frac{1}{v^2} \frac{\partial^2 \Phi}{\partial t^2}=0,
\label{wavequation}
\end{equation}
where $\Phi$ is the angle of twist, $v=\sqrt{G/\rho}$ is the speed of the torsional waves with $G$ the shear modulus and $\rho$ the density of the rod. {\bf To obtain the results for compressional waves, one has to change $\Phi$ by the longitudinal displacement $u$  in the previous equation and use the speed of compressional waves $\sqrt{E/\rho}$ with $E$ the Young modulus}. Since the rod is free at one of its ends ($z=0$), it satisfies the following boundary condition
\begin{equation}\label{Eq:BoundaryCero}
 \left. \frac{\partial \Phi}{\partial z}\right|_{z=0}=0.
\end{equation}
At the other end of the rod ($z=L$), a sinusoidal excitation of intensity $F_0$ and angular frequency $\omega$ is applied, 
\begin{equation}\label{Eq:force}
 \left. \frac{\partial \Phi}{\partial z}\right|_{z=L}= F_0 \exp\left(\rmi\omega t\right),
\end{equation}
where $F_0$ is the ratio between the applied torque and the torsional rigidity~\cite{Graff}. Using separation of variables
\begin{equation}
\Phi(z,t)=\phi(z) \exp\left(\rmi\omega t\right),
\end{equation}
equation (\ref{Eq:linearwaves}) can be written as
\begin{eqnarray}
\frac{\rmd^2 \phi}{\rmd z^2} + \frac{\omega^2}{v^2} \phi = 0,
\end{eqnarray}
with solution
\begin{equation}
\phi(z)=a \exp\left(\rmi k z\right)+ b \exp\left(-\rmi k z\right).
\end{equation}
Here $k=\omega/v$ is the wavenumber and $\phi(z)$ is the time independent angle of twist. The constants $a$ and $b$ can be evaluated as follows. From the boundary condition~(\ref{Eq:BoundaryCero}) one gets
\begin{equation}
\left.\frac{\rmd\phi}{\rmd z}\right|_{z=0}=\rmi k (a-b)=0,
\end{equation}
i.e., $a=b$. Moreover, using~(\ref{Eq:force}) one gets
\begin{equation}\label{Eq:solution}
\phi(z)=-\frac{F_0}{k \sin (kL)}\cos (k z).
\end{equation}
From the last equation it is possible to see that the angle of twist $\phi(z)$ goes to infinity when $\sin(k L)=0$, or well when $k L=n \pi$, $n\in\mathbb{Z}$. This yields an infinite number 
of solutions, $k_n=n \pi/L$, that correspond to the normal mode frequencies 
\begin{equation}\label{Eq:ResonantFreq}
f_n=\frac{nv}{2L},\quad n=1,2,3,\dots
\end{equation}
To avoid the indeterminacy in the response~(\ref{Eq:solution}) it is usual to include some absorption (losses) in a phenomenological way. This can be done by adding an imaginary part to the wavenumber: $k=k_\R+\rmi k_\I$ where $k_\R$ and $k_\I$ are the real and imaginary parts of $k$, respectively. The imaginary part of the wavenumber is a parameter that fix the intensity of the absorption.
{\bf In general $k_\I$ depends on the frequency in a very complicated way~\cite{Crocker} and cannot be taken as a constant for the complete frequency range.} In consequence the angle of twist becomes complex. The response of the rod with absorption is then
\begin{equation}\label{angleoftwist}
\phi(z) = -\frac{F_0}{k}\left[ \frac{\cos(k_\R z)\cosh(k_\I z)-\rmi\sin(k_\R z)\sinh(k_\I z)}
{\sin(k_\R L)\cosh(k_\I L)+\rmi\cos(k_\R L)\sinh(k_\I L)} \right].
\end{equation}
For $z=0$ the intensity of the response is
\begin{equation}\label{ImUz0}
 | \phi(z=0) |^2=\frac{F_{0}^2}{(k_\R^2+k_\I^2)[\sin^2(k_\R L)+\sinh^2(k_\I L)]}.
\end{equation}
Now we will show that equation (\ref{ImUz0}), near to the resonances, can be written in a Breit-Wigner (Lorentzian) form. This regime is called {\em isolated resonances regime}. When $k_\R L=n\pi+\delta$ with $\delta\ll1$, i. e., close to the resonances, and for small absorption, i.e., $k_\I L\ll1$, equation (\ref{ImUz0}) reduces to
\begin{equation}\label{Eq:Lorentzian}
 |\phi(z=0)|^2\approx A_n \frac{\Gamma/2}{\left(f-f_n\right)^2+\left(\Gamma/2\right)^2},
\end{equation}
where $A_n=vF^2_0/2k_\I n^2\pi^3$ and $\Gamma=k_\I v/\pi$. Here, $f_{n}$ and $\Gamma$ are the center and width of the resonance, respectively.

\begin{figure}[!ht]
\centerline{\includegraphics[width=0.95\columnwidth]{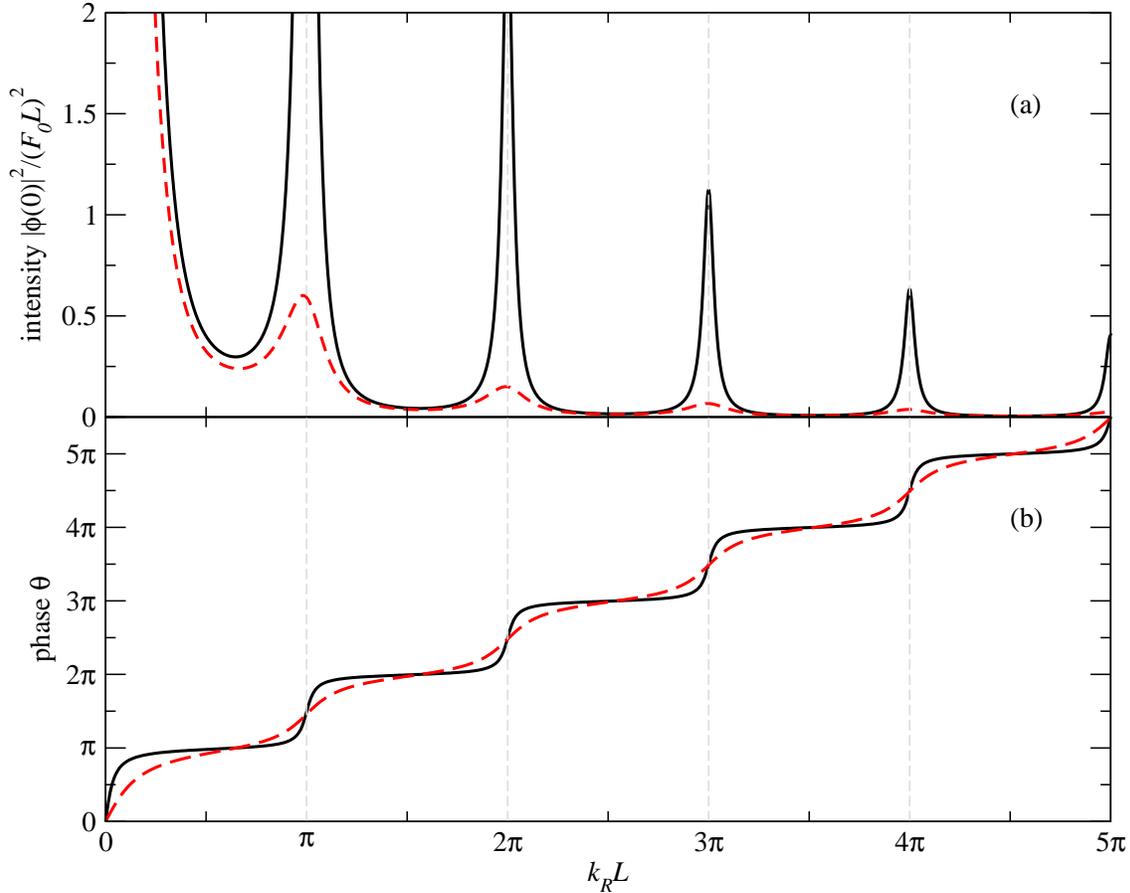}}
\caption{(Color online) a) Intensity $|\phi(z=0)|^2/(F_0 L)^2$ of the angle of twist, equation~(\ref{ImUz0}), and b) phase $\theta$ of the angle of twist, equation~(\ref{Eq:phase}), as a function of $k_\R L$. The continuous (black) lines correspond to  $k_{I}L=0.1$ and the dashed (red) lines correspond to $k_{I}L=0.4$.
}
\label{Fig.ImaginaryPart}
\end{figure}

A plot of $|\phi(z=0)|^2$ as a function of $k_\R L$ is given in figure~\ref{Fig.ImaginaryPart}(a) for two different values of the absorption intensity $k_\I$. 
As can be seen in this figure the intensity of the response shows peaks at $k_\R L=n\pi$, with $n=1,2,3,...$, which correspond to the normal modes. These peaks are usually called resonant. As can be seen in the figure the high of the peaks decreases quadratically as function of $k_\R L$. Additionally one can appreciate in the same figure that the width of the resonant peaks increase with $k_\I$.
{\bf This is only valid when the absorption parameter $k_\I$ is constant and if  $k_\I L \ll 1$; these assumptions are only true locally around the resonant frequencies. Thus, each resonant mode has a different $k_\I$.}

{\bf As we will see in the next section, the detector measures the acceleration of the metal surface. To obtain it, the expression (\ref{angleoftwist}) for the time independent angle of twist $\phi$ should be multiplied by $\omega^2$.
Notice that the Breit-Wigner result for the resonances (\ref{Eq:Lorentzian}),
is still valid since it is assumed that the resonances are narrow
($\delta<<1$) and thus $f\approx f_n$.}

In order to evaluate the phase of the resonance we write the angle of twist in polar form as $\phi(z=0)=|\phi(z=0)|\exp(\rmi\theta)$ where
\begin{equation}\label{Eq:phase}
 \tan\theta=-\frac{k_\R\cos(k_\R L)\sinh(k_\I L)+k_\I\sin(k_\R L)\cosh(k_\I L)}{k_\R\sin(k_\R L)\cosh(k_\I L)-k_\I\cos(k_\R L)\sinh(k_\I L)}.
\end{equation}
A plot of the phase $\theta$ as a function of $k_\R L$ is given in figure~\ref{Fig.ImaginaryPart}(b) for two different values of $k_\I$. Again, for isolated resonances, it is possible obtain a simple expression for the phase:
\begin{equation}\label{Eq:phaseLorentzian}
 \theta\approx\arctan\left(\frac{\Gamma/2}{f-f_{n}}\right).
\end{equation}
From this equation one can see that the phase has a change of $\pi$ for each resonance.

\section{Experimental measurement of resonances in rods}
\label{Sec:Experimental}

To measure the resonances of an aluminum rod we propose to use the experimental setup depicted in figure~\ref{Fig.ExperimentalSetup}. As we will see below, to excite and detect the vibrations, two electromagnetic-acoustic transducers (EMATs) are used~\cite{Morietal,Rossing,Simpson}. The key equipment of the experimental setup of figure~\ref{Fig.ExperimentalSetup} is the vector network analyzer (VNA). A sinusoidal signal is generated by the VNA (ANRITSU model MS4630B) and sent to a high-fidelity Cerwin-Vega! audio amplifier model CV900. The amplified signal is sent to the EMAT exciter located very close to one end ($z=L$) of the rod. The exciter generates a sinusoidal torque on the aluminum rod, see equation~(\ref{Eq:force}), which produces torsional waves of frequency $f$. A second EMAT measures the response at the free end of the rod ($z=0$) and its signal is directly sent to the VNA. Although the range of frequency of the VNA can be swept from $10$~Hz to $300$~MHz, we work in a shorter range since the 
audio amplifier works only from $5$~Hz to $60$~kHz.
{\bf The rod is supported by two nylon threads which are located at the nodes for the lower modes. The effect of this support is very small and decreases considerably for higher modes.
To measure an spectrum, without missing resonances, it is important that both transducers are located close to the ends of the rod since the free boundary conditions guarantee a maximum amplitude of the vibration there.
Notice that moving the detector along the rod will give a measurement of the normal mode wave amplitude.
}

\begin{figure}[!ht]
\centerline{\includegraphics[scale=0.6]{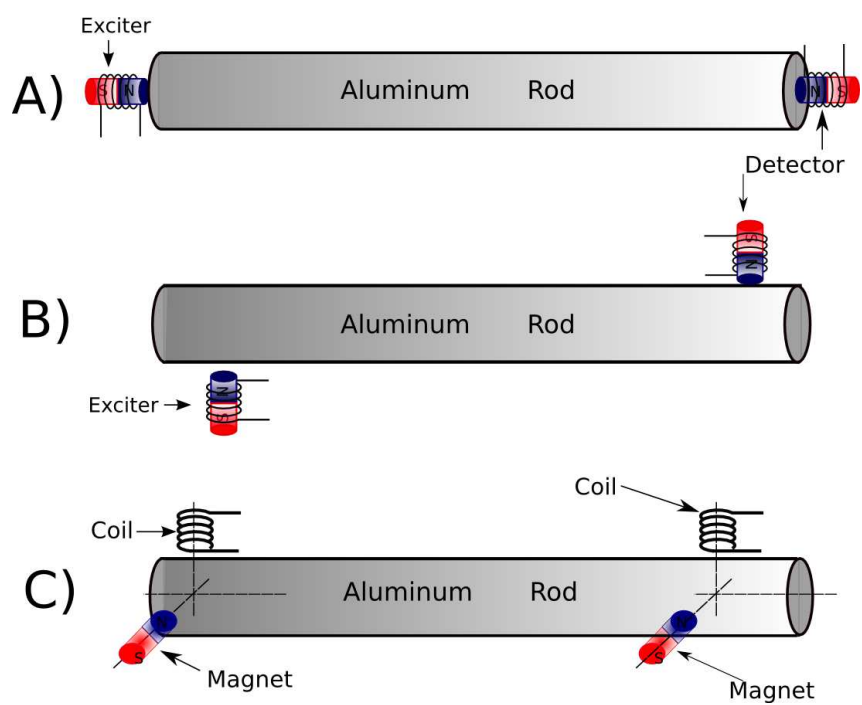}}
\caption{(Color online) EMATs exciter/detector configurations to measure the different kinds of waves in rods: (a) compressional, (b) bending and (c) torsional.}
\label{Fig.EMATSconfig}
\end{figure}

The EMATs can be built easily with coils and permanent magnets; changing the orientation of the coil and the magnet, with respect to the rod axis, different kinds of waves can be selectively excited or detected~\cite{Morietal}. The EMATs are also invertible, i.e., they can be used as exciters or detectors. The configuration of the EMATs to excite (detect) the different kind of waves in rods is shown in figure~\ref{Fig.EMATSconfig}. {\bf Heuristically the EMAT, as an exciter, operates as follows}: a variable current $I(t)$ of frequency $f$ in the coil generates a magnetic field oscillating with the same frequency. When a metal surface is near to the coil, due to Faraday's induction law, eddy currents are produced on the metal. These currents interact with the EMAT's permanent magnet through the Lorentz force. In this way the surface is attracted and repelled at  frequency $f$ without mechanical contact. {\bf Heuristically} the EMAT, as a detector, works as follows: when a vibrating metallic surface is close to the EMAT's permanent magnet, the 
change of the flux of the magnetic field produces eddy currents on the metal {\bf proportional to the speed of the metal}. These currents generate a magnetic field which induce a emf, {\bf proportional to the derivative of these currents}, on the EMAT's coil detector.
{\bf Thus the detector measures the acceleration of the metal surface. Notice that for each normal mode this is proportional to the amplitude of vibration since the frequency is almost constant.}

{\bf In our experiment the EMAT exciter, on the one hand, has a cylindrical neodymium magnet of diameter 12~mm, height 12~mm and 12000~G of residual induction and has a coil with 100 turns and diameter of 40~mm and height of 28~mm. The magnet (enameled) wire was the No. 14~AWG.
The EMAT detector, on the other hand, also has a cylindrical neodymium magnet of diameter 4~mm, height 4~mm and 12000~G of residual induction. A coil with 400 turns of diameter 10~mm and height of 10~mm. 
The magnet wire in this case was the No. 32 AWG. With these characteristics of the EMATs, the power pumped into the EMAT exciter was of the order of 65~W while the typical signal measured with the EMAT detector was of the order
of 200~mV.
}

{\bf The exciter as well
as the detector, have a finite size. Therefore, the interaction with
the rod is not only at one point, as in the theoretical model of
section 2, but in a finite region. This only affects the results when
the wavelength is of the order of the size of the EMATs; this
corresponds to modes with $n\gtrsim 100$; this yields a frequency that
exceeds the maximum operating frequency of the amplifier.}
\begin{figure}[!ht]
\centerline{\includegraphics[width=0.6\columnwidth]{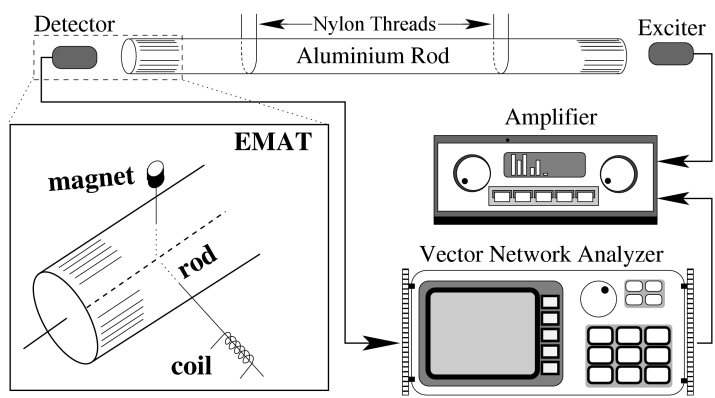}}
\caption{Experimental setup used to measure the resonant response of the aluminum rod. The EMAT (lower left corner) is configured to measure torsional waves. 
{\bf The rod is supported by two nylon threads.}}
\label{Fig.ExperimentalSetup}
\end{figure}

The measurements made with the VNA can be transferred to the computer using the floppy unit or well throughout a direct connection to the computer using the GPIB or RS-232 ports. The VNA allows to measure both, the intensity and the phase of the response.
\begin{figure}[!ht]
\begin{center}
\centerline{\includegraphics[width=0.7\columnwidth]{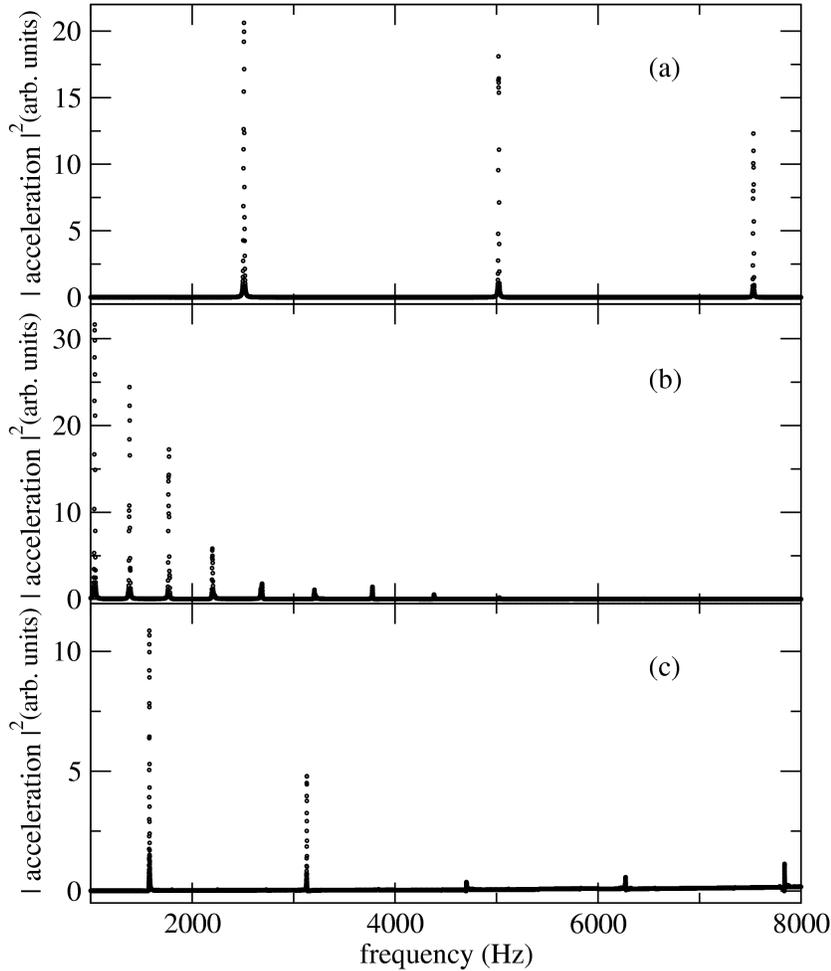}}
\end{center}
\caption{Measured spectrum using the setup of figure~\ref{Fig.ExperimentalSetup} and the EMAT configurations of figure~\ref{Fig.EMATSconfig}. The aluminum rod has a length $L=1$~m and a circular cross section with diameter $D=1.27$~cm. Here (a), (b) and (c) correspond to compressional, bending and torsional spectra, respectively.}
\label{Fig.WideSpectrum}
\end{figure}

\begin{figure}[!ht]
\begin{center}
 \includegraphics[width=0.7\columnwidth]{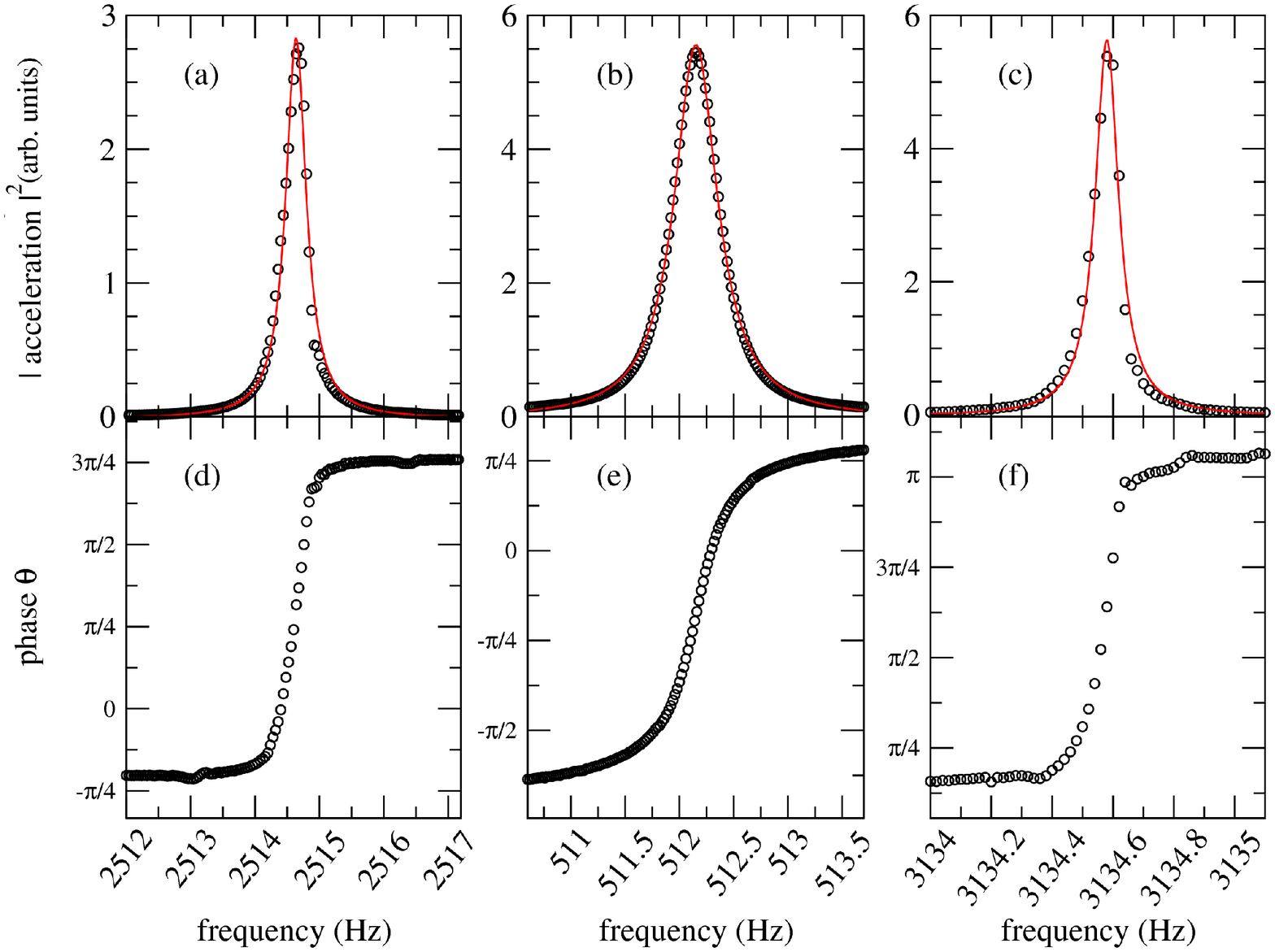}
\end{center}
\caption{(Color online) Measured resonances for the same rod of figure~\ref{Fig.WideSpectrum}.
The upper panels give {\bf the square of the acceleration} and the lower panels the phase $\theta$ of the angle of twist. The solid (red) lines correspond to Lorentzian fits (see equation~\ref{Eq:Lorentzian}) with parameters: (a) compressional 
$\Gamma=0.39$~Hz and $f_{r}=2514.6$~Hz; (b) bending
$\Gamma=0.49$~Hz and $f_{r}=512.2$~Hz; and (c) torsional
$\Gamma=0.09$~Hz and $f_{r}=3134.6$~Hz. Here $f_{r}$ and $\Gamma$ are the center and width of the Lorentzian fit, respectively.}
\label{Fig.ResonantPeaks}
\end{figure}

\section{Comparison Between Theory and Experiment}\label{Sec:Comparison}

\begin{figure}[!ht]
\begin{center}
\includegraphics[width=0.6\columnwidth]{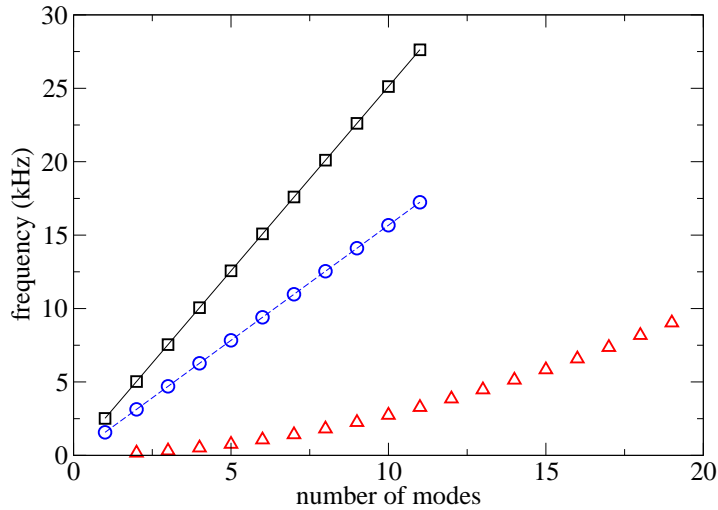}
\end{center}
\caption{(Color online) Resonant frequencies measured with the VNA and the EMATs for the same rod of figure~\ref{Fig.ResonantPeaks} as a function of the number of modes. The results for compressional waves are given by the squares (black), for torsional waves by the circles (blue) and for bending waves by the triangles (red). The solid and dashed lines correspond to a least-squares fit for the compressional and torsional results, respectively.}
\label{Fig.ResonantFrequencies}
\end{figure}

In figure~\ref{Fig.WideSpectrum} we show the measured spectra, with the EMATs in the different configurations of figure~\ref{Fig.EMATSconfig}. As can be seen, the different kind of vibrations are selected by the EMATs since different peaks appear for different configurations. Only a very small compressional peak, around $5$~kHz appears in the bending spectrum. The appearance of this peak is not an incorrect alignment of the coil or magnet, but a real physical effect called lateral inertia correction~\cite{Morietal,Graff}. Three measured peaks, corresponding to compressional, torsional and bending resonances, are given in detail in figure~\ref{Fig.ResonantPeaks}. Apart from {\bf the square of the acceleration (proportional to intensity $|\phi(z=0)|^2$)}, the corresponding phase $\theta(z=0)$ is given in the same figure. Lorentzian fits, see~(\ref{Eq:Lorentzian}), for the resonant peaks are also shown. The absorption parameter $k_\I$ can be obtained from the width of the fitted resonances. As can be seen in figure~\ref{Fig.ResonantPeaks}, the 
compressional and 
bending resonances are wider than the torsional ones. This is expected since the coupling of the torsional wave with the air is very small due to the circular symmetry of the cross-sectional area of the rod and the fact that shear waves cannot travel in air. As expected, the phase of the response shows a change of $\pi$ for each resonance.

The velocity $v$ of the compressional and torsional waves can be obtained measuring several resonances since $f_{n}$ are linear with $n$, and the slope is $v/2L$. In figure~\ref{Fig.ResonantFrequencies} the centers of the resonances as a function of $n$ for torsional, compressional and bending waves are plotted. As can be seen in this figure, the bending waves are not linear with $n$; this means that they are dispersive~\cite{Graff,Landau1}. The slopes of the curves for compressional and torsional waves yield wave speeds of $5\,025\pm5$~m/s and $3\,135\pm1$~m/s, respectively.
{\bf Notice that the uncertainty in the compressional wave velocity is larger than that of the torsional waves, this is due to the Rayleigh or lateral inertia effect (see ref.~\cite{Graff}).}
{\bf Also, from these velocities, some physical parameters of aluminium rod can be calculated. Using a value of $\rho=2722\pm21$~Kg/m$^3$ we obtained Young's modulus $E=68.6\pm0.3$~GPa from the compressional velocity and shear modulus $G=26.7\pm0.2$~GPa.
Those values agree with those found in the literature~\cite{Crandall}.}

\section{Conclusions}\label{Sec:Conclusions}

We have introduced an experiment, for the advanced physics laboratory, which permits undergraduates to learn the basic principles of spectroscopy. It also allows the students to compare experimental results with theoretical predictions since very simple systems, as it is a vibrating rod, can be studied in both ways. The technique, called acoustic resonant spectroscopy, allows the students to measure the resonant curve and the phase of the response. 

The possibilities of this setup in the undergraduate laboratory courses are several since the rod can be substituted by arbitrary and more complicated elastic systems. Apart from being an experiment in which non-destructive testing can be performed, the apparatus can be used with great success to show quantitatively to the students several interesting physical phenomena as the emergence of bands in periodic systems~\cite{JASA1}, the wave amplitudes in plates with regular or irregular shapes~\cite{JFV}, among many others. 
{\bf In fact, the effect of the absorption can also be studied when covering part of the rod with an absorbing foam, the width was found to increase two orders of magnitude.
We should mention that the experimental setup was successfully already used in three advanced physics laboratory courses.} 

\ack

We thank A. Morales and E. Basurto for their help in
the experimental measurements. We also thank X. A. M\'endez-B\'aez, A. Salas-Brito and A. Arreola for useful comments.  This work was supported by DGAPA-UNAM project IN11131 and by CONACYT project 79613.

\end{document}